\documentclass[a4, 11pt]{article}
\usepackage{natbib, amsfonts, amsthm, amsmath,mathpazo,array,graphicx}
\usepackage{comment,color,tikz}
\usetikzlibrary{matrix, positioning}
\tikzset{main node/.style={circle,fill=white,draw,minimum size=1cm,inner sep=0pt},} 
\usepackage{enumitem}
\usepackage[utf8]{inputenc}
\usepackage[english]{babel}
\usepackage{xcolor}
\usetikzlibrary{intersections}
\usepackage{url}
\usepackage{breakurl}
\usepackage[breaklinks=true]{hyperref}
\usepackage{pgfplots}
\usepgfplotslibrary{fillbetween}
\pgfplotsset{compat=1.17}
\usepackage[autostyle, english = american]{csquotes}
\usepackage[hang,flushmargin]{footmisc}
\usepackage{xurl}
\usepackage{color}
\usepackage{nicefrac}
\usepackage{comment} 
\usepackage{tabularx} % for 'tabularx' environment and 'X' column type
\usepackage{booktabs} % for horizontal rules

\title{\vspace{-3.0cm} \hspace{-1.0cm} \Large Measurement challenges in AI catastrophic risk governance and safety frameworks}
\author{}
\author{\small Atoosa Kasirzadeh \\ \small {Carnegie Mellon University} \\ \small (atoosa.kasirzadeh@gmail.com)}
\date{}
\begin{document}
\maketitle

\vspace{-3mm}

\centerline{\fbox{Published in Tech Policy Press (30 September 2024)}\footnote{\url{https://www.techpolicy.press/measurement-challenges-in-ai-catastrophic-risk-governance-and-safety-frameworks/}}}

\vspace{-2mm}

%\linenumbers

\begin{abstract}
\noindent Safety frameworks represent a significant development in AI governance: they are the first type of publicly shared catastrophic risk management framework developed by major AI companies and focus specifically on AI scaling decisions. I identify six critical measurement challenges in their implementation and propose three policy recommendations to improve their validity and reliability.
\end{abstract}

\section*{Introduction}

Safety Frameworks, otherwise known as Scaling Policies, have recently emerged as a critical organizational tool for managing catastrophic risks associated with increasingly capable AI systems. Pioneered by leading AI companies such as Anthropic,\footnote{See \citet{anthropic2023scaling}.} OpenAI,\footnote{See \citet{openai2023preparedness}.} and Google DeepMind,\footnote{See \citet{deepmind2024frontier}.} safety frameworks are created to help AI companies make informed decisions about safely increasing the size and capabilities of AI models. These policies represent an advancement in commitments towards catastrophic AI risk management practices and are notably the first type of risk evaluation framework publicly shared by major AI companies and recommended at the May 2024 Seoul AI summit.\footnote{See \citet{UKGovernment2024FrontierAI}.}

While safety frameworks are a significant step in organizational efforts to govern frontier AI, ensuring the protection of humans and societies from catastrophic AI-related risks requires grounding them in rigorous scientific methodologies. Towards this goal, I argue that measurement modeling as an analytical approach to safety frameworks is essential for enhancing both their operational robustness and scientific validity. 

Measurement modeling, widely used in quantitative and computational social sciences,\footnote{See \citet{jackman2009measurement,bandalos2018measurement,jacobs2021measurement,quinn2010analyze}} offers a critical lens for characterizing and evaluating complex and unobservable theoretical constructs. A measurement model is a statistical model (e.g., a regression model) that connects abstract, unobservable theoretical constructs (e.g., recidivism risk) with observable data (e.g., criminal history, drug involvement). The quality of inferences drawn from these models is typically evaluated using two key concepts: validity (the extent to which a measure accurately reflects the intended construct) and reliability (the consistency of the measurement).

To provide a concrete foundation for this analysis, I primarily focus on Anthropic's safety framework (version 1.0),\footnote{See \citet{anthropic2023scaling}.} which stands as the most comprehensive public document of its kind to date. I then outline how this analysis extends to and informs other safety frameworks. By employing a measurement modeling lens, I identify six neglected problems (See Table~\ref{tab:ai-safety-challenges}) that are crucial to address through collaboration among diverse expert perspectives. First, a collection of models or an embedded model in larger AI ecosystems might trigger catastrophic events. Second, indirect or contributing causes can bring about catastrophic events via complex causal chains. Third, the open-ended characterization of AI Safety Levels (ASLs) can introduce uncertainties into the effective governance of AI catastrophic risks. Fourth, the lack of rigorous justification in setting specific quantitative thresholds presents obstacles in reliably defining and measuring catastrophic events. Fifth, the validity of AI safety assessments can be compromised by the fundamental limitations inherent in red-teaming methodologies. Lastly, mechanisms to ensure developer accountability in ASL classification, particularly for false negatives, are needed to address the risk of AI systems being deployed with inadequate safety measures due to inaccurate catastrophic risk evaluations.

\begin{table}[htbp]
\centering
\begin{tabularx}{\textwidth}{>{\raggedright\arraybackslash}p{0.3\textwidth}>{\raggedright\arraybackslash}X}
\toprule
\textbf{Challenges} & \textbf{Description} \\
\midrule
1. Collective or embedded models impact & Current characterization overlooks catastrophic events triggered by an aggregate effect of multiple AI models or an embedded AI model in a larger social ecosystem. \\
\addlinespace
2. Indirect causation & A narrow focus on "direct causation" may overlook important indirect or contributing factors that could lead to catastrophic events caused by a frontier AI model. \\
\addlinespace
3. Open-ended AI Safety Level (ASL) characterization & Undefined upper bounds for the highest-level ASL-N introduces uncertainties in catastrophic risk assessment. \\
\addlinespace
4. Quantitative threshold justification & The absence of clear and rigorous justification for quantitative thresholds in defining catastrophic events presents significant challenges in reliably identifying and measuring such occurrences. \\
\addlinespace
5. Red-teaming limitations & Fundamental limitations of red-teaming methodologies compromise a guaranteed validity for AI safety evaluation. \\
\addlinespace
6. Developer accountability in ASL classification & Mechanisms are needed to ensure developers could be held responsible for accurate risk assessments in ASL classification, particularly to prevent false negatives. \\
\bottomrule
\end{tabularx}
\caption{Fundamental challenges in catastrophic AI risk governance within safety frameworks}
\label{tab:ai-safety-challenges}
\end{table}

I then propose three policy paths forward: 
\begin{itemize}
\item Substantially revise safety frameworks by addressing challenges 1-6, incorporating diverse expert perspectives on catastrophic events and their causal pathways.
\item Complement current safety frameworks with additional catastrophic risk management frameworks that address challenges 1-6; and
\item Incorporate safety frameworks into robust regulatory structures.
\end{itemize}

\section*{Safety Frameworks}

Safety frameworks offer a structured rubric for making informed decisions on scaling frontier AI models, in order to balance potential benefits against the risks of catastrophic harm. Three notable examples are: Anthropic's Responsible Scaling Policy framework;\footnote{See \citet{anthropic2023scaling}.} OpenAI’s Preparedness framework;\footnote{See \citet{openai2023preparedness}.} and Google DeepMind’s Frontier Safety framework.\footnote{See \citet{deepmind2024frontier}.}

These frameworks share a common goal of systematically evaluating and mitigating potential catastrophic risks associated with the development, deployment, and containment of increasingly powerful AI models. I primarily focus on Anthropic's Responsible Scaling Policy (ARSP) version 1.0 as it is the most comprehensive publicly available document to date, yet the analysis extends to and informs other safety frameworks with similar objectives.

ARSP introduces tiered ASLs to mandate progressively stricter safety and security measures as AI models' capabilities increase. The policy operates on the premise that as AI systems are scaled up, their potential for causing catastrophic events grows proportionally with their capabilities. Consequently, ARSP implements precautionary measures to mitigate the catastrophic risks of AI deployment and containment posed by increasingly capable AI models.

ARSP can be viewed as a measurement model for quantifying the unobservable variable of ``AI catastrophic risk.'' The observable indicators are model capabilities, performance on specific task evaluations (such as Chemical, Biological, Radiological, or Nuclear (CBRN) related tasks), deployment and containment context, and red-team testing results. These indicators are integrated to estimate the level of catastrophic risk, which is then used to classify AI systems into discrete ASLs. The ASL serves as an ordinal representation of the underlying catastrophic risk. The success of this measurement model hinges on its validity and reliability. Validity concerns whether the ASL classifications accurately reflect catastrophic AI risk, while reliability addresses the consistency of these classifications across different evaluators or contexts. 

I analyzed the version 1.0 ARSP from a measurement modeling lens and argue that this version has important limitations that could undermine its goal of preventing catastrophic risks from scaling up AI models. I contend that substantial additional work is necessary to develop an effective safety framework for managing catastrophic risks. Particularly, I examine ARSP across four dimensions and raise challenges to each: defining catastrophic events and AI safety levels, operationalizing these constructs into measurable indicators, reliably ensuring consistent measurement and interpretation of these indicators across different contexts and observers, accountability for when ARSP fails to prevent catastrophic risks from scaled AI models. I then point to directions for improvement.

\subsection*{Construct definition and operationalization}

\textbf{Catastrophic events.} ARSP defines catastrophic events as “large-scale devastation (for example, thousands of deaths or hundreds of billions of dollars in damage) that is directly caused by an AI model and wouldn’t have occurred without it.” \citep[p. 1]{anthropic2023scaling} The policy identifies two main sources of catastrophic risk from frontier AI: misuse and autonomy. Misuse risks involve the intentional use of AI for harmful purposes, such as creating CBRN or cyber threats. Autonomy risks come from AI systems potentially developing increased independence, allowing them to replicate and behave unpredictably, even without deliberate misuse. Two core challenges underpin the scope of this characterization.

\begin{quote}
\emph{Challenge 1. Collective or embedded models might trigger catastrophic events.} 
\end{quote}

ARSP's current definition, which focuses on individual AI models as potential triggers for catastrophic events (see definition above [Anthropic, p. 1]), is becoming increasingly narrow as AI development transitions towards more complex, agentic, and compounded AI ecosystems. The current ARSP framework for characterizing catastrophic events does not sufficiently capture threat scenarios involving multiple interacting AI models or their integration into broader ecosystems. This situation could give rise to the challenge of collective catastrophic risk, where a family of AI models, each individually below the catastrophic risk threshold, might collectively exceed it. Furthermore, the potential for post-deployment enhancement through methods like scaffolding and inference-time computation complicates catastrophic risk assessment. These factors combined make the evaluation of the potential for catastrophic events in evolving AI systems significantly more complex than current frameworks seem to account for.

Two potential scenarios could illustrate this challenge. First, adversaries might exploit combinations of models to achieve harmful outcomes, even when each individual model has been demonstrated to be safe.\footnote{See \citet{jones2024adversaries}.} Second, the cumulative impact of widespread automation across multiple domains could lead to accumulative catastrophic risks, even if no single AI system poses a catastrophic threat on its own.\footnote{See \citet{kasirzadeh2024two}.}

One possible approach to addressing the catastrophic risk of collective or embedded models is the development of richer catastrophe modeling, such as rigorous model interaction simulators. This hypothetical tool would aim to predict potential catastrophic behaviors when multiple AI models interact. Nevertheless, translating the outcomes of such simulations to real-world scenarios could present substantial challenges and constraints. As a result, such a simulator might provide valuable insights, but it should not be relied upon as a sole safeguard.

\begin{quote}
\emph{Challenge 2. Indirect or contributing causes can bring about catastrophic events via complex causal chains.} 
\end{quote}

The current definition of catastrophic events, focusing on "large-scale devastation... directly caused by an AI model," overlooks a critical aspect: indirect causation and salient contributing causes. Indirect causation refers to cases where AI plays a pivotal but not immediately apparent role. For instance, the development and deployment of advanced AI models could trigger an international AI arms race, becoming a salient contributor to increased geopolitical instability or conflict. A concrete example might be AI-enhanced cyber warfare capabilities leading to critical infrastructure failures across multiple countries. 

AI systems might also amplify existing systemic risks or introduce new vulnerabilities that become salient contributing causes to a catastrophic event. The current narrow scope of AI catastrophic events may lead to underestimating the full range of potential catastrophic outcomes associated with advanced AI models, particularly those arising from complex interactions between AI and other sociotechnical systems. This could include scenarios where AI exacerbates climate change through increased energy consumption or where AI-powered misinformation campaigns gradually lead to the breakdown of trust in democratic institutions and social order.

To advance the study of indirect or contributing causes, one option is the creation of a complex causal chain analysis system. This system would leverage tools from causal discovery and inference\footnote{See \citet{spirtes2001causation}.} to simulate complex causal chains leading to AI catastrophes. Nevertheless, while this system could be useful in theory, there is enormous complexity involved in a long-term prediction of indirect catastrophic effects. Such a tool, if feasible, would require rigorous testing, validation, and ongoing refinement to ensure its predictions remain somewhat relevant.

\textbf{AI Safety Levels.} Anthropic's ASL, modeled loosely after the US government's biosafety level (BSL) standards, establishes a structured approach to AI model safety. ASL says if an AI model qualifies for certain dangerous capabilities, it advances to new ASL levels, triggering deployment and containment measures and requirements. ARSP’s ASL levels are defined as follows: ASL1 applies to narrow AI models (e.g., chess players) with no clear catastrophic risk or deployment/containment concerns. ASL2 corresponds to current general purpose AI models with no capabilities likely to cause catastrophe, although early indications may be present (e.g., a general purpose AI system providing unreliable bioweapon-related information). ASL3 represents models that can perform certain advanced dangerous capabilities, either immediately or with minimal additional post-training techniques. ASL4 corresponds to significantly higher risks than ASL3, with capabilities exceeding the best humans in some key areas of concern. ASL-N is provisionally termed for levels beyond ASL4, potentially including systems capable of long-term survival against human resistance or complete automation of scientific research.

\begin{quote}
\emph{Challenge 3. The open-ended characterization of ASLs introduces uncertainties into AI catastrophic risk governance.} 
\end{quote}

The ASL framework's open-ended nature, allowing for undefined upper bounds for the highest-level ASL-N could pose an issue into reliable judgment of catastrophes. Unlike the finite BSL standards (which have a maximum safety level of 4),\footnote{See \citet{CDC2020Biosafety}.} ASL acknowledges AI's unpredictable advancement but introduces substantial uncertainties into risk assessment. The lack of concrete bounded definitions for ASL4 and beyond risks turning safety standards into a moving target, potentially leading to complacency about the dangers of capabilities where safety measures would be postponed to vague future threats. Importantly, this open-ended characterization might create a false sense of safety, implying an ability to indefinitely scale safety measures alongside AI capabilities—an assumption that could prove dangerous if AI progress outpaces the capacity to implement effective safeguards.

Moving forward, we need to engage in a comprehensive and mature dialogue involving AI researchers, ethicists, policymakers, and industry experts to explore the implications of  open-ended safety levels. This discussion should explore whether to maintain this open-endedness or develop methods to establish more reliable upper bounds. The goal is to balance adaptability with clearer guidelines for concretizing safety levels.

\subsection*{Measurement reliability and validity}

\textbf{Catastrophic event measurement.} ARSP establishes quantitative thresholds for "large-scale devastation" in defining catastrophic events, as outlined above.

\begin{quote}
\emph{Challenge 4: The lack of clear and rigorous justification in setting specific quantitative thresholds presents obstacles in reliably defining and measuring catastrophic events.} 
\end{quote}

The introduction of quantitative thresholds for defining AI catastrophic events, such as specific numbers of fatalities or economic damage, raises important questions about methodology and justification. For instance, whether a catastrophic event threshold should be set at 10,000 deaths, 100,000 deaths, or higher requires careful consideration, as does choosing between economic impact benchmarks of hundreds of millions versus hundreds of billions of dollars in damage. These considerations extend beyond mere numbers to questions of authority and power in defining the value of human life.

The ability to set these definitions could significantly influence the establishment of ASL thresholds. The ground for justifications also might need to consider agendas for international cooperation\footnote{See \citet{UKGovernment2023AINetwork}.} as if not, it could lead to conflicts of interest or geopolitical tensions if different entities adopt varying standards. Without clear and rigorous justification, these thresholds may be susceptible to manipulation by stakeholders with misaligned incentives seeking to evade regulatory scrutiny or minimize their AI systems' perceived risks. For example, an entity might argue for a higher fatality threshold to classify their AI system at a lower risk level, thereby avoiding stricter governance.

A critical next step in addressing this challenge is to convene various multidisciplinary, international working groups. These groups should include AI researchers, ethicists, economists, policymakers, and representatives from a variety of expertise backgrounds. Their primary task would be to develop legitimate reasons and arguments for justifying and setting these thresholds.
Key questions to be addressed include: What ethical and economic principles should guide the valuation of human life across different regions? How can we incorporate factors such as societal disruption or catastrophic long-term environmental damage into a definition of AI catastrophic events? How frequently should these thresholds be reviewed and potentially adjusted? The groups should aim to produce a detailed report outlining their recommendations, methodologies, and justifications.

\textbf{ASL measurement.} ARSP mandates regular, conservative evaluations during AI model training and containment to identify risk level increases. These evaluations occur every 4x compute increase and determine if additional safety measures are needed before further scaling. ASL2 involves precursor capabilities to catastrophic risks, requiring model cards, red-teaming, and policy adherence. ASL3 is reached with substantial CBRN risks or low-level autonomous replication abilities. ASL4 and higher safety levels specifics are yet to be determined. ARSP acknowledges that these evaluations are “fundamentally difficult, and there remain disagreements about threat models” [Anthropic, p.7].

\begin{quote}
\emph{Challenge 5: The validity of AI safety assessments could be compromised by the fundamental limitations inherent in red-teaming methodologies.} 
\end{quote}

The reliance on red-teaming as a key evaluation method, while valuable, comes with its own set of fundamental limitations.\footnote{See \citet{feffer2024red}.} The success of red-teaming depends heavily on the expertise and creativity of the testers, which may not always match the potential capabilities of malicious actors or unforeseen edge cases. Moreover, passing a red-team evaluation provides no full guarantee of safety against all possible misuse scenarios. The vast space of potential inputs, contexts, and use cases for AI systems makes it impossible to exhaustively test for all possible vulnerabilities or misuse vectors. The absence of provably safe methods for generative AI models, coupled with a limited understanding of how specific red-teaming results generalize to broader contexts, casts doubt on the ultimate reliability of any safety evaluation protocol primarily grounded in red-teaming efforts. Additionally, the distribution of inputs during deployment might be substantially different from both the training and red-teaming distributions, thus inducing different model behavior and possibly unexpected failures. 

A crucial next step in addressing the limitations of red-teaming methodologies is to develop a more holistic approach to evaluation. This approach could combine traditional red-teaming with formal verification methods, and ongoing real-world monitoring, all tailored to assess the safety implications of scaling AI systems to more advanced capabilities. Until we develop and validate truly robust evaluation techniques, we might seriously consider applying the precautionary principle to higher safety levels. This principle suggests that when an action or policy has a suspected risk of causing severe harm to the public or the environment, in the absence of scientific consensus that the action or policy is safe, the burden of proof that it is not harmful falls on those proposing the action.

Applying the precautionary principle to higher ASLs and safety frameworks would mean at least three things. First, ASL thresholds for higher levels would be set much more conservatively. This means, among other things, creating a larger-than-current safety buffer zone \citet[p. 11]{anthropic2023scaling}, where AI systems would be classified at higher safety levels even with lower estimated risk. This approach ensures stricter oversight and more rigorous safety measures, even if their current risk assessment falls below the typical threshold for that level. Second, before allowing the deployment or scaling of high-level AI systems, extensive and comprehensive testing and validation would be required to demonstrate safety. This might include longer testing periods, more diverse testing scenarios, and higher standards of evidence for safety claims, even if it significantly slows down development or deployment timelines. Third, the implementation of a carefully staged approach to deploying frontier AI systems would allow for incremental testing and risk assessment at each phase of development and rollout. This would involve starting with very limited, tightly controlled environments and gradually expanding the scope and scale of deployment only after thorough safety evaluations at each stage. Each phase would require demonstrable evidence of safety before progressing to the next, potentially extending the overall deployment process but reducing the risk of unforeseen consequences at scale.

\begin{quote}
\emph{Challenge 6: Mechanisms for developer accountability in ASL classification are needed, particularly to address false negatives, which could lead to AI systems with potentially catastrophic consequences being developed and deployed with inadequate safety measures due to inaccurate risk evaluations.}
\end{quote}

A false negative in ASL classification occurs when a model posing higher risks is incorrectly classified at a lower ASL level than appropriate. This accountability problem comes from at least two factors. First, developers using narrow threat models might overlook other potential sources of catastrophic risk, such as systemic risks. Second, developers may evaluate catastrophic capabilities of each model in isolation, failing to address how risks might compound or interact across different AI models or systems, such as agentic instances and interactions with other technologies. 

To address potential accountability challenges in ASL classification, we might consider a two-part accountability approach. (1) Developing a graduated responsibility framework to link financial obligations to the accuracy of catastrophic risk classification, encouraging thorough self-assessments. (2) Incorporating diverse, collaborative oversight panels that could refine classification criteria over time.

\section*{Three policy pathways}

I propose three policy paths forward with safety frameworks given the six challenges identified above and rooted in the provisional next steps for improvement. Full explanation is beyond this piece's scope and will be elaborated in future work.

The \textbf{first path} is to substantially revise safety frameworks by engaging a more diverse expert epistemic background with varying conceptions of catastrophic events, safety cases and levels, and a justified and transparent consensus mechanism. This path can, to a certain extent, address all six challenges identified. It tackles the narrow focus on individual models (challenge 1) and the overlooking of salient contributing causation (challenge 2) by bringing in diverse perspectives that can account for complex AI ecosystems and indirect causal effects. It also addresses the ambiguities in open-ended safety levels (challenge 3) by developing more concrete definitions for higher safety levels. By involving epistemically diverse experts, the measurement criteria for catastrophic events can be established on more rigorous grounds (challenge 4). This path can also enhance the validity of evaluation methods beyond current red-teaming limitations (challenge 5) and improve the accountability in ASL risk classification to reduce false negatives (challenge 6). The outcome would be a substantially improved safety framework addressing the aforementioned challenges.

The \textbf{second path} is to complement the current safety frameworks with additional catastrophic risk management frameworks that incorporate broader risk threat models beyond misuse by bad actors and model autonomy. This path is drawn from the limitations exposed in all challenges. This path suggests that complementary frameworks are needed to address multiple aspects of AI catastrophic risk management: 
\begin{itemize}
\item Measuring collective or bundle model impacts (challenge 1) to account for complex AI ecosystems; 
\item Assessing indirect and salient contributing causes of catastrophic events (challenge 2); 
\item Establishing more precise thresholds and criteria for catastrophic events (challenges 3 and 4); 
\item Enhancing evaluation methods beyond the limitations of current red-teaming approaches (challenge 5); and
\item Improving reliability in risk classification to reduce false negatives (challenge 6). 
\end{itemize}

These additional frameworks would work in conjunction with the ARSP in its current format, creating a more comprehensive approach to AI risk management. Only the combination of these complementary frameworks and the ARSP would be considered fully effective in addressing the multifaceted nature of AI catastrophic risks.

The \textbf{third path} addresses the overarching issues identified across all challenges by integrating safety frameworks into stringent yet flexible regulatory frameworks. This approach recognizes the need for robust oversight while allowing for adaptability, as overly rigid regulations risk becoming rapidly outdated in the face of swift AI advancements. Regulatory oversight could help standardize definitions and thresholds, addressing challenges 2 and 4. It could mandate more comprehensive catastrophic risk assessments, tackling the narrow focus issue identified in challenges 1 and the accountability concern discussed in challenge 6. Furthermore, a regulatory framework could provide clearer guidelines for safety levels and evaluation methods, addressing the concerns raised in challenges 3 and 5. This path recognizes the need for a more standardized framework to AI safety that goes beyond individual company policies.

\textbf{Acknowledgments.} I thank Markus Anderljung, Georg Arndt, Alan Chan, Ben Garfinkel, John Halstead, Leonie Koessler, Aidan O’Gara, Cullen O'Keefe, Sam Manning, Malcolm Murray, Oliver Ritchie, Anna Ringvold, Robert Trager, and Peter Wills for their valuable feedback and insights. I am also grateful to the Centre for the Governance of AI 2024 Winter Fellows for their support and thought-provoking exchanges.

\newpage

\bibliographystyle{chicago}
\bibliography{main}

\end{document}